\documentclass{ws-mpla}
\usepackage{graphicx}

\begin{document}

\markboth{C.~Sasaki, B.~Friman and K.~Redlich}
{Density fluctuations and a 1st-order chiral phase transition}

\catchline{}{}{}{}{}

\title{
Density fluctuations and a first-order chiral phase transition 
\\
in non-equilibrium
}

\author{CHIHIRO SASAKI}

\address{Physik-Department, Technische Universit\"{a}t M\"{u}nchen,
\\
D-85747 Garching, Germany
\\
csasaki@ph.tum.de}

\author{BENGT FRIMAN}

\address{GSI, D-64291 Darmstadt, Germany}

\author{KRZYSZTOF REDLICH}

\address{Institute of Theoretical Physics,
University of Wroclaw,
\\
PL--50204 Wroc\l aw, Poland
\\
and
\\
Institute f\"ur Kernphysik, Technische Universit\"at Darmstadt,
\\
D-64289 Darmstadt, Germany}

\maketitle

\pub{Received (Day Month Year)}{Revised (Day Month Year)}

\begin{abstract}
The thermodynamics of a first-order chiral phase transition 
is considered in the presence of spinodal phase separation using 
the Nambu-Jona-Lasinio model in the mean field approximation. 
We focus on the behavior of conserved charge fluctuations. We show 
that in non-equilibrium the specific heat and charge 
susceptibilities diverge as the system crosses the isothermal spinodal lines.
\end{abstract}

\section{Introduction}	

The search for the critical end point (CEP) is one of the central 
issues in strongly interacting hot/dense QCD matter~\cite{cep,srs}. 
It is of particular interest to identify the position of the CEP
in the phase diagram and to study general properties of thermodynamic 
quantities in its vicinity. 
Modifications in the magnitude of fluctuations or the corresponding 
susceptibilities can be considered as a possible signal for
deconfinement and chiral symmetry 
restoration~\cite{qsus:lattice,lattice:o6,kunihiro,srs,qsus:model}.
In this context, fluctuations related to conserved charges play an 
important role since they are directly accessible in 
experiments~\cite{fluct}.

The enhancement of the baryon number fluctuations could be a clear 
indication for the existence of the CEP in the QCD phase diagram. 
However, finite density fluctuations along the 
first-order transition appear under the assumption that this transition 
happens in equilibrium. This is modified when there is a 
deviation from equilibrium~\cite{our:spinodal}.
In this contribution we briefly show that the enhanced baryon number 
density fluctuations is a signal for the first-order phase transition 
in the presence of spinodal decomposition.

\section{Quark number susceptibility and spinodal instabilities}

In a non-equilibrium system,  a first-order 
phase transition is intimately linked with the existence of a 
convex anomaly in the thermodynamic pressure~\cite{ran}.
There is an interval of energy density or baryon number density 
where the derivative of the pressure, $\partial P/{\partial V}>0$, 
is positive. This anomalous behavior characterizes a region of 
instability in the ($T,n_q)$-plane which is bounded by the spinodal 
lines, where the pressure derivative with respect to volume vanishes. 
The derivative taken at constant temperature and that taken at 
constant entropy,
\begin{equation}
\left( \frac{\partial P}{\partial V} \right)_T=0 
\qquad{\rm and}\qquad 
\left(
\frac{\partial P}{\partial V} \right)_S=0\,,
\end{equation}
define the isothermal and isentropic spinodal lines respectively.

If the first-order phase transition takes place in 
equilibrium, there is a coexistence region, which ends at the CEP.
However, in a non-equilibrium first-order phase transition, 
the system supercools/superheats and, if driven sufficiently far 
from equilibrium, it becomes unstable due to the convex anomaly 
in the thermodynamic pressure. In other words, in the coexistence 
region there is a range of densities and temperatures, bounded 
by the spinodal lines, where the spatially uniform system is 
mechanically unstable.
Spinodal decomposition is thought to play a dominant role in the 
dynamics of low energy nuclear collisions in the regime of the 
first-order nuclear liquid-gas transition~\cite{ran,heiselberg}. 
Furthermore, the consequences of spinodal decomposition
in the connection with the chiral and deconfinement phase transitions 
in heavy ion collisions have been 
discussed~\cite{ran,heiselberg,spinodal:fluc}.

\begin{figure}
\begin{center}
\includegraphics[width=5.5cm]{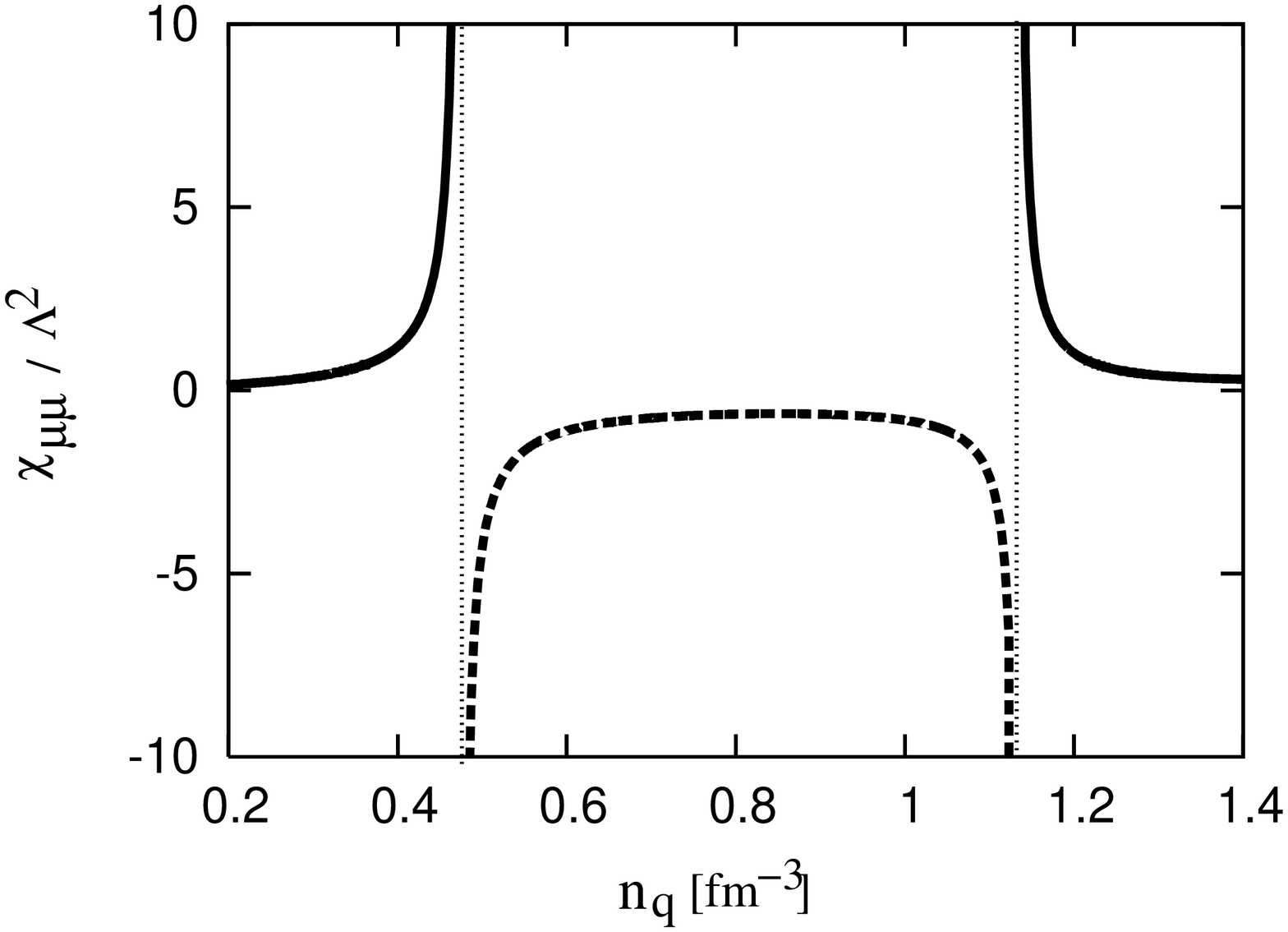}
\includegraphics[width=6.5cm]{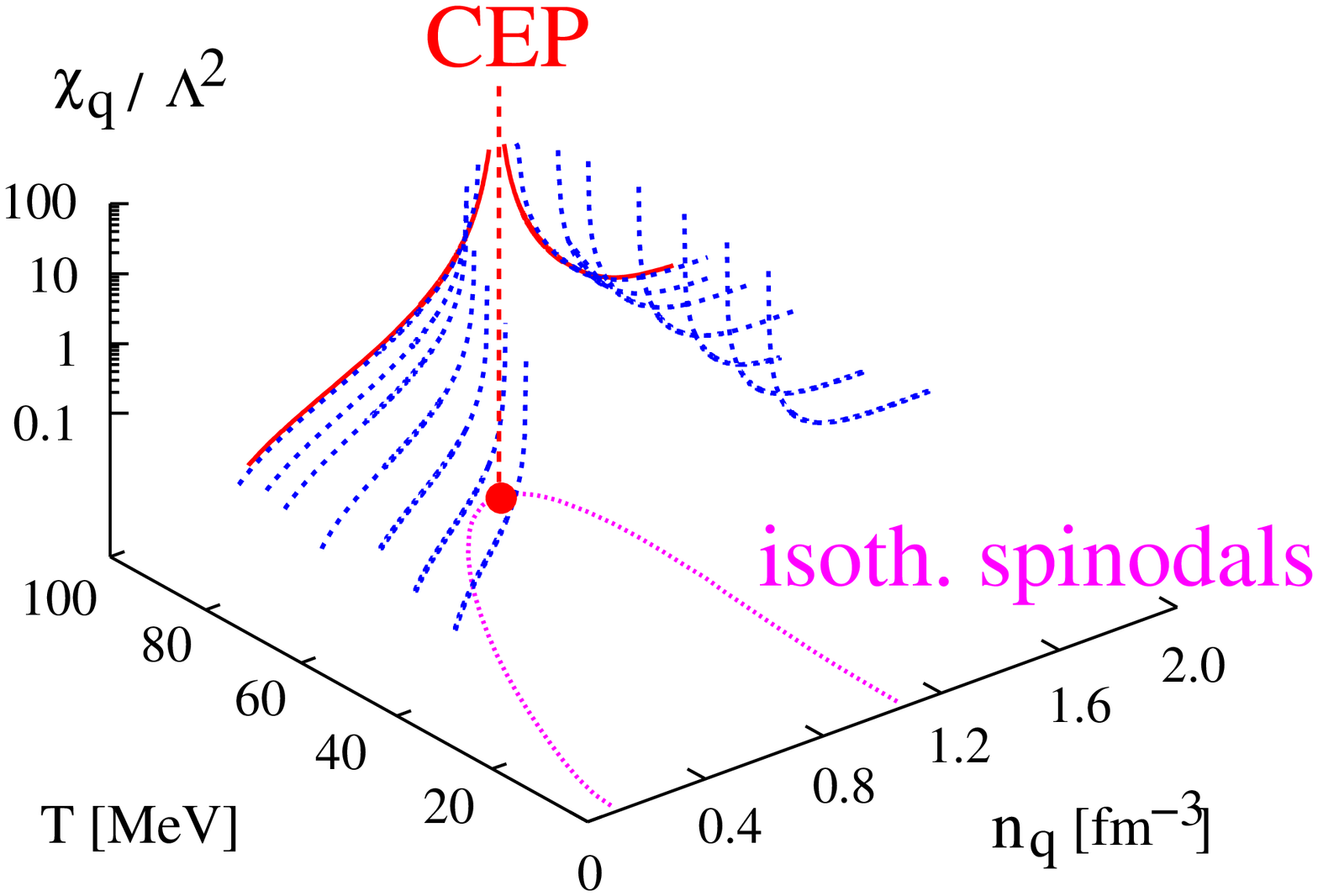}
\caption{
(Left) The net quark number susceptibility at $T=50$ MeV as a
function of the quark number density across the first-order
phase transition. (Right) The net quark number susceptibility
in the stable and meta-stable regions.
}
\label{sus_sp}
\end{center}
\end{figure}

In Fig.~\ref{sus_sp}-left we show the evolution of the net quark 
number fluctuations along a path of fixed $T=50$ MeV calculated 
in the Nambu--Jona-Lasinio (NJL) model in the mean field 
approximation~\cite{njl}.
When entering the coexistence region, there is a 
singularity in $\chi_q$ that appears when crossing the isothermal
spinodal lines and  where the fluctuations diverge and the 
susceptibility changes sign. 
In between the spinodal lines, the susceptibility is negative. 
Consequently, this implies  instabilities in the baryon number 
fluctuations when crossing from a meta-stable to an unstable phase.
The above  behavior of $\chi_q$ is a direct consequence of the 
thermodynamics relation
\begin{equation}
\left( \frac{\partial P}{\partial V} \right)_T
= - \frac{n_q^2}{V}\frac{1}{\chi_q}\,.
\label{pder}
\end{equation}
Along the isothermal spinodals the  pressure derivative  in 
Eq.~(\ref{pder}) vanishes.
Thus, for non-vanishing density $n_q$, $\chi_q$ must diverge to 
satisfy (\ref{pder}).
Furthermore, since the pressure derivative 
${\partial P}/{\partial V}|_T$ changes sign when crossing the 
spinodal line, there must be a corresponding sign change in $\chi_q$,
as seen in Fig.~\ref{sus_sp}-left. 

In Fig.~\ref{sus_sp}-right we show the evolution of the singularity
at the spinodal lines when approaching  the CEP. The critical 
exponent at the isothermal spinodal line is found to be $\gamma=1/2$, 
with $\chi_q \sim (\mu-\mu_c)^{-\gamma}$, while $\gamma=2/3$ at the 
CEP~\cite{our:spinodal}. Thus, the singularities at the two spinodal 
lines conspire to yield a somewhat stronger divergence as they join 
at the CEP. The critical region of enhanced susceptibility around 
the TCP/CEP is fairly small~\cite{schaefer-wambach,our:njl}, 
while in the more realistic non-equilibrium system one expects 
fluctuations in a larger region of the phase diagram, i.e., 
over a broader range of beam energies, due to the spinodal instabilities.

The rate of change in entropy with respect to temperature at constant
pressure gives the specific heat expressed as
\begin{equation}
C_P
= T \left( \frac{\partial S}{\partial T} \right)_P
= TV \left[ \chi_{TT} - \frac{2 s}{n_q}\chi_{\mu T}
{}+ \left( \frac{s}{n_q} \right)^2 \chi_q \right]\,.
\end{equation}
The entropy $\chi_{TT}$ and mixed $\chi_{\mu T}$ susceptibilities 
exhibit the same behaviors as that of $\chi_q$ shown in 
Fig.~\ref{sus_sp}-left. Thus $C_P$ also diverges on the isothermal 
spinodal lines and becomes negative in the mixed phase~\footnote{
 The specific heat with constant volume, on the other hand,
 continuously changes with $n_q$ and has no singularities
 on the mean-field level.
}. It was  argued that in low energy nuclear collisions the negative 
specific heat could be a signal of the liquid-gas phase 
transition~\cite{chomaz}. Its occurrence has recently been reported 
as the first experimental evidence for such an anomalous
behavior~\cite{experiment}.

\section{Conclusions}

We have shown that in the presence of spinodal instabilities 
the net quark number fluctuations diverge at the isothermal 
spinodal lines of the first-order chiral phase transition. 
As the system crosses this line, it becomes unstable with respect 
to spinodal decomposition. The unstable region is in principle 
reachable in non-equilibrium systems that is most likely created 
in heavy ion collisions. Consequently,  large fluctuations of 
baryon and electric charge  densities are expected not only at 
the CEP but also when system crosses  a non-equilibrium 
first-order transition.

\section*{Acknowledgments}

C.S. acknowledges partial support by DFG cluster of excellence
``Origin and Structure of the Universe''.
K.R. acknowledges partial support of the Polish Ministry of National
Education (MENiSW) and DFG under the "Mercator program".

\end{document}